\documentclass[11pt,a4paper]{article}
\pdfoutput=1

\usepackage{jheppub}

\title{\boldmath Divergences and Boundary Modes in N=8 Supergravity}

\author{Finn Larsen and}
\author{Pedro Lisb\~{a}o }

\affiliation{Department of Physics and Michigan Center for Theoretical Physics, University of Michigan,\\
450 Church Street, Ann Arbor, MI 48109-1020, USA}

\emailAdd{larsenf@umich.edu}
\emailAdd{plisbao@umich.edu}

\abstract{
We reconsider the one loop divergence of ${\cal N}=8$ supergravity in four dimensions. 
We compute the finite effective potential of ${\cal N}=8$ anti-deSitter supergravity and interpret it as logarithmic running of the cosmological constant. We find that quantum inequivalence between fields that are classically dual is due to boundary modes in AdS$_4$. Some subtleties are traced to the difference between the Euler characteristic of global and thermal AdS$_4$. 
}

\begin{document}
\maketitle
\flushbottom

\section{Introduction and Summary}
\label{sec:introduction}

Divergences in quantum gravity are famously severe and suggestive that long distance physics depends sensitively on the shortest lengths. Supersymmetry mitigates the divergences so effectively that for maximal  ${\cal N}=8$ supergravity in four asymptotically flat dimensions it has not yet been established what divergences remain, if any \cite{Bern:2006kd, Bern:2007xj, Kallosh:2008mq, BjerrumBohr:2009zz, Kallosh:2009db, Bjornsson:2010wm, Dixon:2010gz}. On the other hand, it has long been known that in curved backgrounds, highly relevant for gravity, even the one-loop vacuum amplitude diverges \cite{Christensen:1978gi, Christensen:1978md, Christensen:1979iy, Sezgin:1980tp, Bastianelli:2005vk, Bastianelli:2005uy}. 
The apparent incompatibility between these results created controversy already in the 1980's \cite{Siegel:1980jj, Duff:1980qv, Siegel:1980ax, Gibbons:1984dg, Inami:1984vp, Fradkin:1984ai, Buchbinder:2008jf}. In this paper we revisit this tension from a modern perspective informed by the AdS/CFT correspondence \cite{Maldacena:1997re}.

To exhibit the central issue in more detail it is convenient to focus on the anomalous contribution to the 
trace of the energy momentum tensor
\begin{equation}
\label{eq:emtrace}
\langle T_\mu^\mu \rangle_{\rm an}= \frac{1}{(4\pi)^2} \left( c W^2 - a E_4\right)~,
\end{equation}
where the square of the Weyl tensor $W^2={\rm Riem}^2 - 2{\rm Ric}^2 + \frac{1}{3}R^2$ and the Euler density $E_4={\rm Riem}^2 - 4{\rm Ric}^2 + R^2$ encode dependence on the background geometry\footnote{We assume for simplicity a renormalization scheme where other possible terms are absent.}. The coefficients $c,a$ depend on the matter content of the theory and they have been studied in great detail; e.g. using perturbation theory in small curvature around flat space. Their values for fields with simple couplings to the background have long been established and are summarized in table \ref{table:centralcharges} (later). These well known coefficients are such that, for the field content of ${\cal N}=8$ supergravity, their sum does not vanish. This fact establishes a divergence that is present already at one loop. 

However, there are equally well established perturbative nonrenormalization theorems based on the helicity supertraces over the on-shell spectrum
\begin{equation}
\label{eq:supertrace}
\sum (-)^{2h} h^n=0~,
\end{equation}
for $n<{\cal N}=8$. These sum rules imply powerful cancellations for perturbative amplitudes in asymptotically flat space and related supertrace formulae are influential in particle phenomenology because they survive spontaneous breaking of supersymmetry. For us the important point is that the helicity sum rules establish one-loop nonrenormalization in ${\cal N}=8$ AdS$_4$ supergravity (gauged ${\cal N}=8$ supergravity) \cite{Gibbons:1984dg, Inami:1984vp, Nicolai:1983me}. These cancellations even generalize to all massive levels obtained from Kaluza-Klein compactification of ${\cal N}=1$ supergravity in $11$ dimensions. 

We will argue that despite appearances there is no contradiction, but rather a topological distinction encoded in the boundary conditions. The basis for the sum rules (\ref{eq:supertrace}) is Lorentzian AdS$_4$ which, after Euclidean continuation, gives rise to $S^1\times S^2$ boundary conditions with the $S^1$ corresponding to Euclidean time. In this geometry the Euler characteristic
\begin{equation}
\label{eq:gaussbonnet}
\chi =\frac{ 1}{ 32\pi^2} \int E_4 + ~{\rm bndy} ~,
\end{equation}
vanishes. This is significant because the divergences uncovered by the curvature expansion are proportional to $\chi$ and so they are not captured by AdS$_4$ with $S^1\times S^2$ boundary conditions. On the other hand, we will easily reproduce them from Euclidean AdS$_4$ with $S^3$ boundary conditions since this geometry has Euler invariant $\chi=1$. 

One might wonder if these divergences have any physical significance. We argue in the affirmative by 
computing a finite and nonvanishing one-loop correction to the cosmological constant in maximal 
AdS$_4$ supergravity. In this computation it is manifest that the helicity supertrace 
relations (\ref{eq:supertrace}) are violated in spacetime with $S^3$ boundary conditions. Interestingly, the violation is rather mild so all power law corrections in fact cancel. Thus the cosmological constant acquires just logarithmic running. This feature is intriguing since it might offer a mechanism that could describe dark energy without sacrificing naturalness.  

Our results are subject to an important subtlety that was noticed already in early studies of quantum fields in curved space: quantum inequivalence \cite{Duff:1980qv}. In our context an important example is the relation between a massless antisymmetric tensor and a scalar field. In the classical theory they are equivalent by a field redefinition but their quantum partition functions are related by a shift
%
%
that is proportional to the Euler characteristic (\ref{eq:gaussbonnet}) \cite{Bastianelli:2005vk, Bastianelli:2005uy}. The coefficient of the logarithmic divergence we study therefore depends on the duality frame which becomes part of the data that defines the theory. We interpret this feature as a genuine physical effect: antisymmetric tensor fields support boundary modes that have no analogues in the corresponding scalar field theory. 

In this paper we primarily interpret ${\cal N}=8$ AdS$_4$ supergravity as a low energy effective field theory in its 
own right but ultimately the UV completion involves the full string/M-theory. As an intermediate step we consider 
the theory as compactification of 11D supergravity on AdS$_4\times S^7$. This procedure defines a preferred duality frame for the massless fields and it happens that it is precisely the frame where all logarithmic divergences cancel \cite{Duff:1980qv}. In this setting boundary modes cancel divergences.  

There have been many other recent studies of quantum corrections to AdS spaces in various dimensions. A basic feature of this research is that divergences remain even when supersymmetry is maximal and those divergences 
are related to effects that are unambiguously physical in the dual theory. Some examples: 
\begin{itemize}
\item
In AdS$_{d+1}$ with odd $(d+1)$ there are bulk divergences interpreted as finite quantum anomalies in the dual theory with even $d$. For example, in the case of $d=4$ such anomalies are responsible for the shift $N^2\to N^2-1$ that is expected and confirmed in ${\cal N}=4$ SYM with $SU(N)$ gauge groups \cite{Mansfield:2003gs, Beccaria:2014xda}. 

\item
Quantum corrections to higher spin theories in AdS provide impressive evidence for higher spin holography. 
\cite{Giombi:2013fka, Giombi:2014iua, Giombi:2014yra}

\item
The Bekenstein-Hawking area law for black holes is subject to $\log A$ corrections with coefficients determined by the low energy theory. For BPS black holes these coefficients are determined by divergences in AdS$_2$ and AdS$_2\times S^2$ which are generically nonvanishing (including for ${\cal N}=8$), and their values are confirmed by the microscopic theory in cases where the latter has been established. \cite{Banerjee:2010qc, Banerjee:2011jp, Sen:2011ba, Sen:2012cj, Bhattacharyya:2012wz, Sen:2012dw, Sen:2014aja, Keeler:2014bra, Larsen:2014bqa}
\end{itemize}
Our study of AdS$_4$ was motivated in part by these and related developments. Computations in these contexts share the techniques we employ and offer some confidence in their applicability.


\section{One Loop Quantum Corrections in AdS$_4$}

In this section we employ heat kernel methods to compute the one loop contributions to the anomalous trace of the energy momentum tensor in AdS$_4$ from fields with various spins. We interpret the resulting divergences in the effective action as logarithmic running of the effective cosmological constant. 

\subsection{Notation and Review}

One loop quantum corrections in Euclidean quantum gravity are determined by a Gaussian path integral
with the schematic form,
\begin{equation}
W=-\ln \int\mathcal{D}\phi e^{-\phi \Box \phi} = \frac{1}{2}\ln \det \Box = \frac{1}{2}\sum_i \ln \lambda_i~,
\end{equation}
where the $\phi$ denotes the collection of linearized fields, $\Box$ represents their kinetic operator, and $\lambda_i$ are the eigenvalues of $\Box$. We represent the effective action $W$ in terms of the heat kernel $D(t) = \sum_i e^{-t\lambda_i}$ as
\begin{equation}
\label{eq:Wdef}
W = -\int_{\epsilon^2}^{\infty}\frac{dt}{2t} D(t)~,
\end{equation}
where $\epsilon$ is a UV regulator with dimension of length. It is customary to express results for heat kernels in terms of the (equal point) heat kernel density $K(t)$ expanded at small $t$ 
\begin{equation}
\label{eq:Kdef}
K(t) =  \frac{1}{\textrm{Vol}_{\textrm{AdS}_4}} D(t)=  \frac{1}{(4\pi t)^2}\bigg( 1 + a_2 t + a_4 t^2+ ...\bigg)~.
\end{equation}
Departures from the flat space limit are encoded in the two derivative correction $a_2$ proportional to the Ricci scalar and the four derivative correction $a_4$ that is a linear combination of Riemann squared, Ricci squared, and Ricci scalar squared.\footnote{The volume diverges, since AdS$_4$ is noncompact. We mostly consider local quantities in a homogeneous space so any regulator will apply. The standard renormalized value $\textrm{Vol}_{\textrm{AdS}_4} = {4\pi^2\ell^4_A\over 3}$ will appear later from global considerations with explicit boundary terms.}

We divide the one loop effective action (\ref{eq:Wdef}) into divergent contributions
\begin{equation}
\label{eq:Wdiv}
W_{\textrm{div}} =  \frac{1}{32\pi^2} \left( - {1\over 2\epsilon^4} - a_2 {1\over\epsilon^2} + a_4 \ln \epsilon^2\right) \textrm{Vol}_{\textrm{AdS}_4}~,
\end{equation} 
and a remainder that is finite. From either piece we can form the trace of the energy momentum tensor
\begin{equation}
T^{\mu}_{\mu} = \frac{2}{\sqrt{-g}}g^{\mu\nu}\frac{\delta W}{\delta g^{\mu\nu}}~.
\end{equation}
The logarithmic divergence of the effective action (\ref{eq:Wdiv}) gives an anomalous contribution that is conventionally presented as
\begin{equation}
\label{eq:acdef}
\langle T_\mu^\mu \rangle_{\rm an}= \frac{1}{(4\pi)^2} a_4 = \frac{1}{(4\pi)^2} \left( c W^2 - a E_4\right)~.
\end{equation}
In the nonconformal theories we consider there may be additional contributions to the trace of the energy momentum tensor. 

The values of $c$ and $a$ have been computed perturbatively by many researchers using different methods and schemes \cite{Christensen:1978md, Eguchi:1980jx, Birrell:1982, Vassilevich:2003xt}. The values that are now standard (up to caveats discussed later in this section) are summarized in table \ref{table:centralcharges} below.
\bgroup
\def\arraystretch{1.5}
\begin{center}
\begin{table}[h]
\begin{tabular}{|c|c|c|c|}
\hline
\textbf{Field} & $c$ & $a$ & $c-a$\\ \hline
Real Scalar & $\frac{1}{120}$ & $\frac{1}{360}$ &$\frac{1}{180}$\\ \hline
Weyl Fermion & $\frac{1}{40}$ & $\frac{11}{720}$&$\frac{7}{720}$  \\ \hline
Vector & $\frac{1}{10}$ & $\frac{31}{180}$&-$\frac{13}{180}$ \\ \hline
Antisymmetric Tensor & $\frac{1}{120}$ & $-\frac{179}{360}$&$\frac{91}{180}$ \\ \hline
Gravitino & $-\frac{411}{360}$ & $-\frac{589}{720}$ &$-\frac{233}{720}$\\ \hline
Graviton & $\frac{783}{180}$ & $\frac{571}{180}$&$\frac{53}{45}$ \\ \hline
\end{tabular}
\caption{Central charges $c$ and $a$ for minimally coupled massless fields. Each entry is a physical field with two degrees of freedom except the scalar, which has just one degree of freedom.}
\label{table:centralcharges}
\end{table}
\end{center}
\egroup

\subsection{Computations in AdS$_4$}
We now revisit these computations in the context of AdS$_4$. This geometry is conformally flat so the Weyl tensor vanishes and therefore the central charge $c$ plays no role. Our focus on $a$ is complementary to techniques that impose Einstein's equations in vacuum and identify just the Riemann-squared terms which have coefficient $c-a$. 

The natural representations for fields in AdS$_4$ are the symmetric, transverse, and traceless (STT) tensors with spin $s$. The heat kernels for these fields were comprehensively analyzed by Camporesi and Higuchi \cite{Camporesi:1992wn, Camporesi:1994ga, Higuchi:1986wu} (and recently developed further \cite{Giombi:2014yra}) both using explicit mode functions and also using group theory. We present their results for 
the AdS$_4$ heat kernel of a massive spin $s$ field with conformal dimension $\Delta$ as
\begin{equation}
\label{eq:KAmaster}
K^{(s,\nu)}(t) = \frac{1}{\ell_A^4}\int_0^\infty d\lambda ~\mu_s(\lambda) ~e^{-\frac{t}{\ell_A^2} (\lambda^2+\nu^2)}~,
\end{equation}
where $\nu^2 = (\Delta - \frac{3}{2})^2$. The conformal dimension $\Delta$ is equivalent to the mass of the field and in the context of AdS$_4$ it is $\Delta$ that provides the simplest representation of this parameter. Crucially, the Plancherel measure $\mu_s(\lambda)$ for the integration over the continuous eigenvalues $\lambda$ is different for bosons\footnote{We 
simplify notation by absorbing a numerical factor in the Plancherel measure.}
\begin{equation}
\mu_ s(\lambda) = (s+{1\over 2}) {\lambda^2 +(s +{1\over2})^2 \over 4\pi^2}  \lambda \tanh(\pi \lambda)~,
\end{equation}
and for fermions 
\begin{equation}
\mu_ s(\lambda) = (s+{1\over 2}) {\lambda^2 +(s +{1\over2})^2 \over 4\pi^2}  \lambda \coth(\pi \lambda)~.
\end{equation}

The distinction between bosons and fermions is inconsequential in the UV region where $\lambda\to\infty$ since then both $\tanh(\pi \lambda)\to 1$ and $\coth(\pi \lambda)\to 1$. It is instructive to evaluate the heat kernel (\ref{eq:KAmaster}) such that this common feature is manifest. For bosons we write $\tanh(\pi \lambda) = 1- \frac{2}{e^{2\pi\lambda}+1}$ and then find 
\begin{align}
\label{eq:boseheatkernel}
K^{(s,\nu)}_{\rm boson}(t) &= \frac{s+{1\over 2}}{4 \pi^2\ell^4_A}e^{-{\frac{t\nu^2}{\ell_A^2}}} \left[ \int_0^\infty e^{-\frac{t\lambda^2}{\ell_A^2}}(\lambda^2 +(s+\frac{1}{2})^2)\lambda d\lambda
- 2 \int_0^\infty e^{-\frac{t\lambda^2}{\ell_A^2}} \frac{\lambda^2 +(s+\frac{1}{2})^2}{e^{2\pi \lambda}+1}\lambda d\lambda\right] \nonumber \\ 
& =  {s+\frac{1}{2}\over 8\pi^2\ell^4_A} e^{-\frac{t\nu^2}{\ell^2_A}}\left( {\ell^4_A\over t^2} + {\ell^2_A\over t}(s+{1\over 2})^2\right)
- {s+\frac{1}{2}\over 8\pi^2\ell^4_A}\left( {7\over 480} + {(s+{1\over 2})^2\over 12}\right) \nonumber \\
& = \frac{s+{1\over 2}}{8\pi^2\ell^4_A}\left[ {\ell_A^4\over t^2} + {\ell_A^2\over t}\left((s+{1\over 2})^2 - \nu^2\right)\right] + 
\frac{s+\frac{1}{2}}{16\pi^2\ell^4_A}\bigg[ \nu^4 - (s+\frac{1}{2})^2(2\nu^2+\frac{1}{6})-\frac{7}{240}\bigg]~.
\end{align}
The first integral contains the UV terms that are common to bosons and fermions and is elementary for all $t$. The second integral is special to bosons. It is finite for small $t$ so we evaluate it at $t=0$, omitting higher powers in $t$. It is evident from this structure that only the first integral contributes to the terms that are divergent in the UV limit $t\to 0$. 

We next compare with the fermion heat kernel where we write 
$\coth(\pi \lambda) = 1 + \frac{2}{e^{2\pi\lambda}-1}$ and find
\begin{align}
\label{eq:fermiheatkernel}
K^{(s,\nu)}_{\rm fermion}(t) &= \frac{s+{1\over 2}}{4 \pi^2\ell^4_A}e^{-{\frac{t\nu^2}{\ell_A^2}}} \left[ \int_0^\infty e^{-\frac{t\lambda^2}{\ell_A^2}}(\lambda^2 +(s+\frac{1}{2})^2)\lambda d\lambda
+ 2 \int_0^\infty e^{-\frac{t\lambda^2}{\ell_A^2}} \frac{\lambda^2 +(s+\frac{1}{2})^2}{e^{2\pi \lambda}-1}\lambda d\lambda\right]  \nonumber \\
& = \frac{s+{1\over 2}}{8\pi^2\ell^4_A}\left[ {\ell_A^4\over t^2} + {\ell_A^2\over t}\left((s+{1\over 2})^2 - \nu^2\right)\right] + 
\frac{s+\frac{1}{2}}{16\pi^2\ell^4_A}\bigg[ \nu^4 - (s+\frac{1}{2})^2(2\nu^2-\frac{1}{3})+\frac{1}{30}\bigg]~.
\end{align}
Since the first integral is the same in the boson and fermion heat kernels (\ref{eq:boseheatkernel}, \ref{eq:fermiheatkernel}) these expressions have the same divergences in the UV limit $t\to 0$. It is for the same reason that they have the same dependence on conformal dimension appearing through $\nu^2 = (\Delta - \frac{3}{2})^2$. However, the two cases are of course different due to the second integral and this is reflected in the terms that are constant and independent of $\nu$. 

We are particularly interested in massless particles since those are the ones that appear in standard ${\cal N}=8$ supergravity. In AdS$_4$ masslessness is not well characterized by the absence of a mass term in the Lagrangian but rather by the reducibility of the field representation. Representations at spin $s$ generally have dimension $2s+1$ but some special ones are reducible and allow decoupling of a ghost representation that has spin $s_{\rm ghost}=s-1$ and so dimension $2s_{\rm ghost}+1=2s-1$. This leaves two physical degrees of freedom for massless particles with spin, as expected. Group theory methods show that this reduction is possible precisely when the conformal dimension is $\Delta = s+1$ (and so $\nu = \Delta-{3\over 2} = s -\frac{1}{2}$) and also specify that the spin $s-1$ ghosts have $\Delta_{\rm ghost}=s+2$ \cite{Metsaev:1994ys}. These results do not strictly apply for the lowest spins $s={1\over 2}, 0$ but we can apply them formally with the understanding that the ghost subtraction in fact enhances a real scalar to a complex representation.\footnote{For spin $s={1\over 2}$ the rule formally subtracts ghosts that have spin $s_{\rm ghost}=-{1\over 2}$ but that is inconsequential since this representation has dimension $2s_{\rm ghost}+1=0$. For spin $s=0$ it formally subtracts a ghost with spin $s_{\rm ghost}=-1$ and dimension $2s_{\rm ghost}+1=-1$ which effectively adds one degree of freedom, turning one boson into two, with conformal dimensions $\Delta=1,2$.} These rules give 
\begin{equation}
\label{eq:Kbosemassless}
K^{(s, {\rm massless})}_{\rm boson}(t) = K^{(s, s+1)}_{\rm boson}(t) - K^{(s-1, s+2)}_{\rm boson}(t) = \frac{1}{16\pi^2\ell^4_A}\bigg(\frac{2\ell_A^4}{t^2} + \frac{8s^2 \ell_A^2}{t}  -5s^4 + s^2 - \frac{2}{15}\bigg)~,
\end{equation}
for a massless boson with spin $s$, and 
\begin{equation}
\label{eq:Kfermimassless}
K^{(s, {\rm massless})}_{\rm fermion} = (-)\left[ K^{(s, s+1)}_{\rm fermion}(t) - K^{(s-1, s+2)}_{\rm fermion}(t) \right]= \frac{1}{16\pi^2\ell^4_A}\bigg( -\frac{2\ell_A^4}{t^2} - \frac{8s^2 \ell_A^2}{t}+ 5s^4 - \frac{5}{2}s^2 - \frac{13}{240}\bigg)~,
\end{equation}
for a massless fermion with spin $s$. We inserted a sign for the fermion by hand in order to take statistics into account. 

The $t=0$ poles in the massless heat kernels are the same for bosons and fermions (up to the sign that was inserted for fermions) as we expected since that is the case for each of the underlying massive representations. On the other hand, some of the terms that are finite as $t\to 0$ differ, also as expected. This feature is the origin of the apparent lack of pattern in the heat kernel coefficients that is evident when we consider the finite parts of $K_{\textrm{massless}}$ for the first few spins in table \ref{table:Kmassless}. 
\bgroup
\def\arraystretch{1.5}
\begin{center}
\begin{table}[h]
\begin{tabular}{|c|c|c|}
\hline
\textbf{Spin} &$ 16\pi^2 \ell^4_A K^{\rm finite}_{\textrm{massless}}$ & $a$  \\ \hline
$0$ & $-\frac{2}{15}$ & $\frac{1}{180}$ \\ \hline
$\frac{1}{2}$ & $-\frac{11}{30}$ & $\frac{11}{720}$ \\ \hline
$1$ & $-\frac{62}{15}$ & $\frac{31}{180}$ \\ \hline
$\frac{3}{2}$ & $\frac{589}{30}$ & $-\frac{589}{720}$ \\ \hline
$2$ & $-\frac{1142}{15}$ & $\frac{571}{180}$ \\ \hline
\end{tabular}
\caption{The values of $K_{\textrm{massless}}$ computed in AdS$_4$ and the corresponding $a$ anomalies. All entries including the scalar $s=0$ refers to two degrees of freedom.}
\label{table:Kmassless}
\end{table}
\end{center}
\egroup

Our results for the finite parts of the heat kernel $K(t)$ in AdS$_4$ are identical to the $a_4$ coefficients introduced in (\ref{eq:Kdef}) up to a factor $(4\pi)^2$. It can be further recast in terms of the $a$-anomaly introduced in 
(\ref{eq:acdef}) by noting that the Gauss-Bonnet density is $E_4 = 24/\ell^4_A$. We have included the  $a$-anomaly computed this way in table \ref{table:Kmassless}. These values agree perfectly with the results from the local expansion in curvature summarized in table \ref{table:centralcharges}. 

There is a caveat to this agreement. As we have stressed, our computation (which in fact closely follows Camporesi and Higuchi \cite{Camporesi:1993mz}) determines the $a$-anomaly unambiguously for all spin. In contrast, many researchers compute both $c$ and $a$ for low spin but results for $s={3\over 2}, 2$ (and above) are not widely quoted and there is no obvious consensus on their values. This situation is tied with the background dependence of the linearized equations of motion for such fields. The most secure data points are for $c-a$ which is defined in Ricci flat backgrounds and $a$ which, as we have stressed, is unambiguous in maximally symmetric spacetimes. For $s={3\over 2}, 2$ the values of $a, c$  given in table \ref{table:centralcharges} were obtained by combining the results for $a$ given in table \ref{table:Kmassless} with the standard values of $c-a$. 

\subsection{Extended SUSY}
The $t$-poles in the heat kernels (\ref{eq:Kbosemassless}) and (\ref{eq:Kfermimassless}) correspond to 
power law divergences in the effective action. The boson and fermion contributions to these divergenes cancel when 
\begin{equation}
\label{eq:ssumrule}
\sum (-)^{2s} s^n = 0~,
\end{equation}
for $n=0,2$. The massless spectrum only comprises maximal helicity where $|h|=s$ so this condition is equivalent to 
the helicity sum rule (\ref{eq:supertrace}) for $n=0,2$. This is satisfied for ${\cal N}\geq 3$ supergravity and we will focus on these theories. 

\bgroup
\def\arraystretch{1.5}
\begin{center}
\begin{table}[h]
\begin{tabular}{|c|c|c|}
\hline
\textbf{Spin} & \textbf{Conformal Dimension} $\Delta$ & $SO(8)$ \textbf{Multiplicity} \\ \hline
$2$ & $3$ & $1$ \\ \hline
$\frac{3}{2}$ & $\frac{5}{2}$ & $8$ \\ \hline
$1$ & $2$ & $28$ \\ \hline
$\frac{1}{2}$ & $\frac{3}{2}$ & $56$ \\ \hline
$ 0$& $1$ & $35$ \\ \hline
$0$ & $2$ & $35$ \\ \hline
\end{tabular}
\caption{The conformal dimensions and multiplicities of the massless multiplet in $\mathcal{N}=8$ supergravity.}
\label{table:conformaldimensions}
\end{table}
\end{center}
\egroup

For maximal $\mathcal{N}=8$ SUGRA the standard spectrum given in table
\ref{table:conformaldimensions} satisfies the sum rule (\ref{eq:ssumrule}) even for $n=4, 6$ yet the sum of the boson and fermion heat kernels do not vanish
\begin{equation}
\label{eq:kn8}
K_{{\cal N}=8}^{\textrm{total}} =  \langle T_\mu^{\mu}\rangle_{\textrm{ren}} = \frac{1}{16\pi^2\ell_A^4}(-60)~.
\end{equation}
This is possible because the bosonic and fermionic heat kernels (\ref{eq:Kbosemassless}-\ref{eq:Kfermimassless}) are given different polynomials in the spin $s$. 

We can represent the heat kernel result (\ref{eq:kn8}) for ${\cal N}=8$ supergravity as an $a$ anomaly for the entire multiplet,
\begin{equation}
\label{eq:N=8div}
a_{{\cal N}=8} = \frac{5}{2}~.
\end{equation}
Considering also the values of $c$ from table \ref{table:centralcharges} we find that the central charge $c=0$ for the full ${\cal N}=8$ multiplet. We collect these results in table \ref{table:centralcharges8}.

The quantum effective action can be computed from the heat kernel (or, equivalently, from the trace of the energy momentum tensor) by the integral (\ref{eq:Wdef}). We perform the integration with the dimensionless conformal weights $\Delta$ kept fixed. This is justified by the boundary perspective where the dual theory is conformal in the leading approximation and also from the bulk point of view where all fields are in the massless representations that do not even exist for other values of the conformal weights. Since we focus on theories with no power law 
corrections the integrand is a constant and, with the measure indicated in (\ref{eq:Wdef}), the integral gives a logarithmically divergent term in the effective action. 

Multiple research groups have reported that in fact the trace anomaly does vanish for ${\cal N}=8$ supergravity in AdS$_4$ \cite{Gibbons:1984dg, Inami:1984vp} and so there are no divergences. Those results refer to different boundary conditions where the spectrum is discrete and the helicity sum rule (\ref{eq:supertrace}) applies for all $n<{\cal N}$. We will return to this in more detail in the next section. 

\bgroup
\def\arraystretch{1.5}
\begin{center}
\begin{table}[h]
\begin{tabular}{|c|c|c|c|}
\hline
& $c$ & $a$ & $c-a$\\ \hline
Massless $\mathcal{N}=8$ multiplet & $0$ & $\frac{5}{2}$ &$-\frac{5}{2}$\\ \hline
\end{tabular}
\caption{Central charges $c$ and $a$ for ${\cal N}=8$ supergravity.}
\label{table:centralcharges8}
\end{table}
\end{center}
\egroup

\subsection{Interpretation of Quantum Corrections}

The anomalous contribution to the trace of the energy momentum tensor is independent of position because spacetime is homogeneous. A classical cosmological constant in the action similarly gives a constant contribution but the origin of the anomalous contribution is a divergence $W_{\rm div} = {1\over 2} D_0\ln \epsilon^2/\ell^2_0$ in the effective action that manifests itself in the renormalized action as a term that evolves logarithmically
\begin{equation}
W_{\rm ren} = -  {1\over 2}D_0\ln {x^2_{\rm phys}\over\ell^2_0}~.
\end{equation}
The renormalization scale $\ell_0$ enters as an IR cutoff on the integral over the heat kernel. It is arbitrary but of order of the AdS-scale. The physical length scale $x_{\rm phys}$ depends on the process as usual and may be anywhere in the range from much smaller than the AdS scale (for UV processes) to much larger than the AdS scale (for the IR properties). 

We interpret the scale dependent quantum effective action as a contribution $\delta\Lambda$ to the cosmological constant determined by
\begin{equation}
W_{\rm ren}  = - {{\textrm{Vol}_{\textrm{AdS}_4}}\delta\Lambda \over 8\pi G}~. 
\end{equation}
It is convenient to express the running in terms of the effective AdS scale $\ell_{\rm eff} = \sqrt{- 3/\Lambda}$:
\begin{equation}
\label{eq:lrunning}
{1\over\ell^2_{\rm eff}} = {1\over\ell^2_A} \left[ 1 - {4\pi G\over 3\ell^2_A} (K_0\ell^4_A)\ln {x^2_{\rm phys}\over\ell^2_0}\right] ~.
\end{equation}
The combination $(K_0\ell^4_A)$ is a pure number that we have computed above for some specific fields. The most important part of this expression is the absence of power law corrections that would enter through the UV cutoff $\epsilon$. This would signal dependence on unknown UV physics. Instead we have nontrivial logarithmic quantum corrections that are computable within the low energy theory. \footnote{Logarithmic running of the cosmological constant was discussed also in \cite{Taylor:1989ua, Bytsenko:1994at}.}

A good way to construct AdS supergravity is to gauge supergravity in flat space. This procedure identifies the 
gauge coupling constant as \cite{Freedman:1976aw, deWit:1981yv}
\begin{equation}
e^2 = {4\pi G\over\ell^2_A}~. 
\end{equation}
This coupling constant is small $e^2\ll 1$ when the AdS radius is much larger than Planck scale as we have implicitly presumed. Resumming the (possibly large) logarithms we can recast (\ref{eq:lrunning}) as
\begin{equation}
e^2 = {e^2_0\over 1 + {1\over 3}e^2_0 (K_0\ell^4_A)\ln {x^2_{\rm phys}\over\ell^2_0}} ~.
\end{equation}
Comparing with standard formulae we can write an effective $\beta$-function for these theories
\begin{equation}
\beta = - {b\over 16\pi^2} e^3~.
\end{equation}
where
\begin{equation}
b = - {1\over 3} (16\pi^2K_0\ell^4_A)~. 
\end{equation}
The $\beta$-function determines the running of a dimensionless version of the cosmological constant 
through the usual renormalization group equations. The numerical coefficient $b=8a$ is $b=20$ for ${\cal N}=8$ supergravity, $b=8(1 + n_V/4)$ for ${\cal N}=4$ supergravity with $n_V$ matter multiplets, and similarly for other examples. 

Our computations are all made in bulk and there is no reference to a boundary theory. This is a rather old fashioned point of view but it is worthwhile for interpreting the set up as a toy model for the physical cosmological constant. For this we imagine the signs such that the cosmological constant is positive and the running such that it becomes small at large distances. The dimensionless coupling $e^2$ would be tuned to take a tiny value, of order $10^{-120}$. The absence of power law corrections would then ensure naturalness in the sense that the logarithmic running is so mild that quantum corrections would preserve the enormous hierarchy. This mechanism does not explain the smallness of the observed cosmological constant but it offers a viable scenario for its naturalness.

\section{Quantum Inequivalence and Boundary Modes.}

In this section we discuss the interplay between the trace of the energy momentum tensor and quantum inequivalence between duality frames. We interpret quantum inequivalence as a physical effect due to boundary modes. We also show that the divergences and the boundary modes are both related to the topology of global AdS$_4$. 

\subsection{Quantum Inequivalence.}
A classical antisymmetric tensor in four dimensions can be mapped into a scalar field via the duality transformation
\begin{equation}
\label{eq:Bphiduality}
H_{\mu\nu\sigma} = 3\nabla_{[\mu}B_{\nu\sigma]} = \epsilon_{\mu\nu\sigma\lambda}\nabla^\lambda \phi~.
\end{equation}
These fields are therefore classically equivalent. However, one loop corrections in curved space do not respect this equivalence. For example, the trace anomaly coefficients for these two fields differ as displayed in table \ref{table:centralchargesAs}. This leads to the conclusion that these fields are quantum inequivalent \cite{Duff:1980qv}.

\bgroup
\def\arraystretch{1.5}
\begin{center}
\begin{table}[h]
\begin{tabular}{|c|c|c|c|}
\hline
& $c$ & $a$ & $c-a$\\ \hline
Antisymmetric Tensor & $\frac{1}{120}$ & $-\frac{179}{360}$&$\frac{91}{180}$ \\ \hline
Real Scalar & $\frac{1}{120}$ & $\frac{1}{360}$ &$\frac{1}{180}$\\ \hline
${\rm As} - \phi$ & $0$ & -$\frac{1}{2}$ &$\frac{1}{2}$\\ \hline
\end{tabular}
\caption{Central charges $c$ and $a$ for the $2$-form, the real scalar, and their evanescent difference.}
\label{table:centralchargesAs}
\end{table}
\end{center}
\egroup

However, in some sense the dual fields do not differ by terribly much. They have identical physical spectra as captured by propagating on-shell degrees of freedom: the ``evanescent'' field defined by their difference has no propagating degrees of freedom. Although the $a$-anomaly coefficients do indeed differ, the $c$-anomaly coefficients do not; and the $a$-anomaly is the coefficient of the Euler density which is topological. Many researchers therefore argue that these fields are equivalent, at least for all practical purposes \cite{Grisaru:1984vk, Fradkin:1984ai, Buchbinder:2008jf}. 

Our discussion of divergences in ${\cal N}=8$ supergravity (and related theories) is intertwined with quantum inequivalence. First of all, the divergence (\ref{eq:N=8div}) is entirely an $a$-anomaly, the $c$-anomaly of ${\cal N}=8$ supergravity vanishes. We nevertheless interpret this divergence physically in terms of the logarithmic evolution of the cosmological constant. This assigns physical significance to the $a$-anomaly even though it has a  topological aspect. 

Next, the value of the $a$-anomaly, and therefore its physical significance, depends on the duality frame. Concretely, one might choose to dualize any number of antisymmetric tensors into scalars, or vice versa, affecting the trace of the energy momentum tensor in the process. Therefore such dualizations are not symmetries.

\subsection{AdS$_4$ SUGRA from 11D.}
The default spectrum of ${\cal N}=8$ supergravity summarized in table \ref{table:conformaldimensions} comprises $70$ scalars and no antisymmetric tensors. Comparing tables \ref{table:centralcharges8} and \ref{table:centralchargesAs} we find that a duality frame where exactly five scalars are represented instead as antisymmetric tensors exhibits no trace anomaly. 

It turns out that this precise number is a natural expectation when approaching supergravity in AdS$_4$ as compactification of 11D supergravity on $S^7$. The 11D 3-form with components $a_{IJK}$ is reduced into various lower forms in 4D including 3-forms and 2-forms,
\begin{align}
a_{\mu\nu\sigma}(x,y) &=b_{\mu\nu\sigma}(x) Y(y)~, \\ \nonumber
a_{\mu \nu p}(x,y) &=b_{\mu \nu}(x) Y_p^{(CE)}(y) + \tilde{b}_{\mu \nu}(x) Y_p^{(E)}(y)~.
\end{align}
The AdS$_4$ coordinates are denoted by $x$ and greek indices, while their $S^7$ counterparts are $y$ coordinates and latin indices. The functions $Y(y)$, $Y_p^{(CE)}(y), Y_p^{(E)}(y)$ are spherical harmonics on $S^7$ that are respectively a scalar, a coexact 1-form, and an exact 1-form. 

The $2$-tensor $\tilde{b}_{\mu \nu}(x)$ is the coefficient of $Y^{(E)}(y) = dY(y)$ which is effectively a scalar on $S^7$ so there is one of these modes, while $b_{\mu \nu}(x)$ is the coefficient of $Y^{(CE)}(y) = {}^*dY(y)$ which is effectively a transverse vector on $S^7$ with six modes. Thus there is a total of $1+6=7$ $2$-tensors in the effective 4D theory as one would also expect from toroidal compactification of 11D supergravity to 4D. Classically these seven antisymmetric tensors can be dualized to seven scalars but in view of quantum inequivalence this must be done with care. 

The $3$-form tensor $b_{\mu\nu\sigma}(x)$ is the coefficient of the ordinary spherical harmonic so there is just one of these fields in four dimensions. A massless $3$-form has no propagating degrees of freedom in four dimensions since the classical equations of motion force it to be constant. At the quantum level gauge fixing of the $3$-form gives two $2$-form ghosts with fermi statistics, three $1$-form ghosts with bose statistics, and four scalar ghosts with fermion statistics. This counting gives $ 4 - 2\cdot 6 + 3\cdot 4 - 4\cdot 1=0$ net components and so no propagating degrees of freedom, as expected. However, as we repeatedly stress, $2$-forms must be handled with care at the quantum level and that applies also to the two ghosts that accompany the $3$-form tensor. At the quantum level one $3$-form tensor contributes with $(-2)$ $2$-forms that cannot be naively dualized to scalars. 

In summary, the duality frame that arises naturally through the AdS$_4$ compactification of 11D supergravity on $S^7$ gives a net of five antisymmetric 2-tensors:
\begin{equation}
1 + 6 - 2= 5~.
\end{equation}
In this duality frame the trace of the energy momentum tensor vanishes and there are no divergences \cite{Duff:1980qv}. 

This result does not invalidate our claim that there are divergences in ${\cal N}=8$ supergravity. On the contrary, it implicitly confirms the notion that different duality frames are quantum inequivalent since otherwise the distinction between $2$-forms and scalars would be meaningless and there would be no utility in counting $2$-forms arising from Kaluza-Klein reduction of 11D supergravity. From the low energy effective field theory point of view it is legitimate to consider AdS$_4$ supergravity with a different number of $2$-forms, including none at all, although it must be understood that such theories might not arise in string theory \cite{Green:2007zzb} and they could be vulnerable to some subtle quantum inconsistency. 

In this paper we focus on massless fields in 4D but the computations can be generalized to the full KK tower of massive fields. All these contributions are again proportional to the Gauss-Bonnet invariant and, level by level, they are nonvanishing. One may sum over all KK towers and recast the remaining divergences in 11D where they become power law divergences. They generally appear at the four derivative order but in the duality frame favored by 11D supergravity they only appear at six derivative order. However, eleven is odd and in odd dimensions all these divergences are nonuniversal and scheme dependent so it is not clear that they are physical. The divergence that is definitely physical is again a logarithm which is due to zero-modes of the two form gauge symmetry. These zero modes were understood from the 11D perspective \cite{Bhattacharyya:2012ye} and the resulting logarithmic correction agrees with the one expected from the solution of the dual ABJM theory via localization \cite{Marino:2009jd, Fuji:2011km, Marino:2011eh}. We hope to report on these elaborations elsewhere.

\subsection{Boundary Modes in AdS$_4$}
The evanescent part of the $2$-form  --- the quantum contribution of an antisymmetric tensor that is above and beyond that of its dual scalar field --- is naturally interpreted as a boundary mode, at least in the context of AdS$_4$. A boundary mode is formally a pure gauge field configuration but it is physical because the putative gauge parameter is non normalizable and so the field configuration cannot be gauged away by any element of the symmetry group. This mechanism is unimportant in classical field theory but it matters in the quantum theory, as expected for a feature related to quantum inequivalence. 

The boundary modes reside in the kernel of the classical duality transformation (\ref{eq:Bphiduality}) between an antisymmetric tensor and a scalar. Their $3$-form field strength vanishes identically in bulk, since they are formally pure gauge, and so they are not assigned a scalar dual
\begin{equation}
H^{\textrm{(bndy~mode)}}_{\mu\nu\sigma} = 0 = \epsilon_{\mu\nu\sigma\lambda}\nabla^\lambda \phi~,
\end{equation}
since a constant scalar $\phi$ is not normalizable in noncompact spacetimes. This is the source of quantum inequivalence from our point of view.

{\it A priori} any field with gauge symmetry might possess one or more boundary modes. For example, in global AdS$_2$ all fields with a gauge symmetry have them \cite{Larsen:2014bqa, Camporesi:1994ga, Sen:2011ba}. On the other hand, in AdS$_{d+1}$ with higher $d$ it was found by explicit construction in global AdS$_{d+1}$ that boundary modes exist only for $\frac{d+1}{2}$-forms \cite{Camporesi:0000, Bhattacharyya:2012ye}. In AdS$_4$ those are precisely the $2$-forms that we are interested in. 

To make the discussion explicit we write the background AdS$_4$ metric
\begin{equation}
\label{eq:globads}
ds^2_4 = \ell^2_A\left(d\rho^2 + \sinh^2\rho d\Omega^2_3\right)~.
\end{equation}
We take the AdS$_4$ radius $\ell_A=1$ in the remainder of this section to avoid cluttered formulae. The normalized boundary modes in this background are 
\begin{align}
\label{eq:boundarybmodes}
B_{\rho i} &= \sqrt{\frac{k+1}{2}}\frac{1}{\sinh\rho}\tanh^{k+1}(\rho/2)\Theta^{(k,\sigma)}_i(\Omega_3)~, \\ \nonumber
B_{i j } & = \sqrt{\frac{1}{2(k+1)}}\tanh^{k+1}(\rho/2) [\tilde{\nabla}_ i \Theta^{(k,\sigma)}_j(\Omega_3) - \tilde{\nabla}_ j \Theta^{(k,\sigma)}_i(\Omega_3)]~.
\end{align}
for $k=1,2,...$. The covariant derivative $\tilde{\nabla}$ refers to components along $S^3$ and latin indices represent these angular components. The $1$-form field $\Theta^{(k,\sigma)}_i(\Omega_3)$ is a vector spherical harmonic with eigenvalue of the Hodge de Rham operator $=(k+1)^2$. The quantum numbers $k,\sigma$ are analogous to the numbers $lm$ used for scalar harmonics on $S^2$ but for vector harmonics $k=0$ is excluded. 

The antisymmetric $2$-form with components (\ref{eq:boundarybmodes}) can be represented as pure gauge $B=dA$ where the $1$-form potential $A$ has components
\begin{align}
\label{eq:boundarypotential}
A_\rho &=0~, \\ \nonumber
A_i &= \frac{1}{\sqrt{2(k+1)}}\tanh^{k+1}(\rho/2)\Theta^{(k,\sigma)}_i(\Omega)~.
\end{align}
This $1$-form does not have finite norm
\begin{align}
\int \sqrt{g}|A|^2 dV &= \int  \sinh^{3}\rho( A^*_{\rho}A_{\rho} g^{\rho \rho} + A^*_{ j}A_{l}g^{jl}) d\rho d\Omega \cr
 & \propto \int_0^\infty   \sinh \rho \tanh^{2k+2}(\rho/2) d\rho = \infty~.
\end{align} 
The inverse metric $g^{jl}$ contributes with a factor of $\sinh^{-2}(\rho)$ that dampens the radial integral at large $\rho$, but insufficiently to render it finite. However, the tensor $B=dA$ is normalizable for all $k=1,2,\ldots$. 
\begin{align}
\label{eq:boundarynorm}
\int \sqrt{g}|B|^2 dV &= \int  \sinh^{3}\rho (2 B^*_{\rho i}B_{\rho j} g^{\rho \rho} g^{ij} + B^*_{i j}B_{lk}g^{ik}g^{jl}) d\rho d\Omega \cr
 & \propto \int_0^\infty   \sinh^{-1}(\rho) \tanh^{2k+2}(\rho/2) d\rho < \infty~.
\end{align} 
The index structure here gives enough factors of the inverse metric $g^{jl}$, contributing each with $\sinh^{-2}(\rho)$, such that their product with the field components is sufficient to overcome the volume factor. The normalization in (\ref{eq:boundarybmodes}) was chosen so that the integral (\ref{eq:boundarynorm}) is unity. The $2$-tensor has support in bulk but we interpret it as a boundary mode because it is locally pure gauge. 

Once we have identified a $1$-form $A$ that gives rise to a $2$-form boundary mode $B=dA$ we should note that gauge equivalent $1$-forms $A'=A+d\Lambda$ give rise to the same boundary mode. The boundary modes thus belong to the two-form cohomology. In order to not overcount them we must impose a gauge condition, taken in (\ref{eq:boundarypotential}) as $A_\rho=0$. 

In summary: while the $2$-form modes (\ref{eq:boundarybmodes}) are formally pure gauge they are physical because the would-be gauge function is non normalizable. Therefore, they contribute to the quantum path integral. Moreover, we have argued that unlike all other modes of the massless $2$-form field, the boundary modes are not captured by the scalar dual. We focus on the massless case for clarity but the quantum inequivalence between a massive $2$-form and its (classically) dual massive vector is similarly due to boundary modes for the $2$-form. 

\subsection{Counting Boundary Modes}
We can find the contribution of the boundary modes to the heat kernel and related quantities by explicitly counting modes, following \cite{Camporesi:0000, Bhattacharyya:2012ye}. The wave function of each mode is normalized to unity so the total number of modes is
\begin{equation}
\label{eq:numbermodes}
n_{\textrm{bndy~modes}} = \sum_{\textrm{all~modes}} \int d^4x \sqrt{g} |B|^2~.
\end{equation}
The sum in equation (\ref{eq:numbermodes}) is over the family of modes presented in (\ref{eq:boundarybmodes}) that is parametrized by the quantum numbers $k,\sigma$.
\begin{align}
\label{eq:simplifyB}
 \int d^4x \sqrt{g} \sum |B|^2 &=  \int d^4x \sqrt{g} \sum (2 B^*_{\rho i}B_{\rho j} g^{\rho \rho} g^{ij} + B^*_{i j}B_{lk}g^{ik}g^{jl})~, \\ \nonumber
 &= \int d^4x \sqrt{g} \sum_{k,\sigma} 2 \frac{ k+1}{  2}   \frac{\tanh^{2k+2}(\rho/2) }{ \sinh^4\rho} |\Theta^{(k,\sigma)}_i(\Omega)|^2~,  \\ \nonumber
 &+\int d^4x \sqrt{g} \sum_{k,\sigma}  \frac{1}{  2(k+1)}   \frac{\tanh^{2k+2}(\rho/2) }{ \sinh^4\rho} |\tilde{\nabla}_ i \Theta^{(k,\sigma)}_j(\Omega_3) - \tilde{\nabla}_ j \Theta^{(k,\sigma)}_i(\Omega_3)|^2 ~.
\end{align}
We can simplify this sum using integration by parts on the angular dependence of the second term,
\begin{align}
\int  |\tilde{\nabla}_ i \Theta^{(k,\sigma)}_j - \tilde{\nabla}_ j \Theta^{(k,\sigma)}_i|^2 d\Omega_3 &= -2 \int  \Theta^{j(k,\sigma)*}\tilde{\nabla}^ i (\tilde{\nabla}_ i \Theta^{(k,\sigma)}_j - \tilde{\nabla}_ j \Theta^{(k,\sigma)}_i) d\Omega_3\\ \nonumber
&= 2(k+1)^2 \int \Theta^{j(k,\sigma)*} \Theta^{(k,\sigma)}_jd\Omega_3~.
\end{align}
In the last step we identified the operator acting on  $\Theta^{(k,\sigma)}_j$ as minus the Hodge deRham operator acting on vectors. We insert this result into (\ref{eq:simplifyB}), combining both contributions into one. One could evaluate the sum over modes at any point but, given that AdS$_4$ is homogeneous, it is sufficient to consider the origin $\rho=0$ where only the $k=1$ spherical harmonic contributes,
\begin{align}
\label{eq:rhotozero}
\sum_{\textrm{all~modes}}  |B|^2 &=  \lim_{\rho \rightarrow 0}  \sum_{k=1}\sum_{\sigma} 2(k+1) |\Theta^{(k,\sigma)}_i(\Omega)|^2    \frac{\tanh^{2k+2}(\rho/2) }{ \sinh^4\rho} ~,\\ \nonumber
& = \frac{1}{ 4} \sum_{\sigma}|\Theta^{(1,\sigma)}_i(\Omega)|^2   ~.  
\end{align}
The sum over $|\Theta^{(k,\sigma)}_i(\Omega)|^2$ for fixed $k$ is proportional to the degeneracy of the $S^3$ vector spherical harmonics,
\begin{equation}
\label{eq:sumovers3}
 \sum_{\sigma} \Theta^{(k,\sigma)*i}(\Omega) \Theta^{(k,\sigma)}_i(\Omega)\bigg|_{k=1} = \frac{6}{\textrm{Vol}_{S^3}}= \frac{3}{\pi^2}~,
 \end{equation}
since there are $2k(k+2)=6$ vector spherical harmonics on $S^3$ with $k=1$. Collecting formulae, the number of boundary modes (\ref{eq:numbermodes}) becomes
\begin{equation}
n_{\textrm{bndy~modes}} = \sum_{\textrm{all~modes}} \int d^4x \sqrt{g} |B|^2 = 
 \frac{3}{4\pi^2} \int d^4x \sqrt{g} = 1~.
\end{equation}
We used the standard regulated volume ${\textrm{Vol}_{\textrm{AdS}_4}}={4\pi^2\over 3}$ since then the result looks nice and intuitive. However, in the current context of a noncompact and maximally symmetric space we should really focus on the density of modes. Indeed, the boundary modes have vanishing eigenvalue of the kinetic operator so they formally contribute by the ``number'' $D_0^{(\textrm{bndy})} = n_{\textrm{bndy~modes}}$ to the constant part $D_0$ of the heat kernel $D(t)$ and this corresponds to the heat kernel density 
\begin{align}
K_0^{(\textrm{bndy})} &=  \frac{D_0^{(\textrm{bndy})}}{\textrm{Vol}_{\textrm{AdS}_4}} = \frac{3}{4\pi^2},
\end{align}
independently of the value assigned to ${\textrm{Vol}_{\textrm{AdS}_4}}$. Comparing with the definition of $a_4$ in (\ref{eq:Kdef}) and the introduction of the $c,a$ anomaly coefficients in (\ref{eq:acdef}) we find
 \begin{equation}
 \label{eq:abndy}
a^{(\textrm{bndy})}=-\frac{1}{2}~.
\end{equation}
since the Euler density $E_4 =24$ in AdS$_4$ with unit radius $\ell_A=1$. 

The value (\ref{eq:abndy}) of the boundary anomaly agrees precisely with the $a$ anomaly of the evanescent difference between a massless antisymmetric tensor and a scalar reported in table \ref{table:centralchargesAs}. This quantitative agreement shows that the quantum inequivalence between an antisymmetric tensor and a dual scalar field is due to boundary modes. This in turn establishes a physical distinction between the inequivalent fields.

\subsection{The Gauss-Bonnet Theorem in AdS$_4$}
We have emphasized the divergences that remain in AdS$_4$ even for maximal SUSY and their interrelation with quantum inequivalence, because these aspects are the most interesting to us and they have not been developed in recent literature. Another approach to one-loop effects that is closer aligned with conventional wisdom invokes reflecting boundary conditions on all modes \cite{Avis:1977yn, Breitenlohner:1982jf, Gibbons:1984dg, Inami:1984vp}. This leads to a discrete sum over modes, the helicity sum rule (\ref{eq:supertrace}) applies in full, and there are no divergences at one loop  (and well beyond). The relation between these results involves global aspects
of AdS$_4$, as captured by the Euler invariant. It is therefore instructive to evaluate the Euler invariant in detail. 

The curvature tensor in a maximally symmetric spacetime is constant so the Gauss-Bonnet integral over the Euler density is proportional to the volume
\begin{equation}
\label{eq:bulke4}
\int E_4 = \int {\rm Tr} ~{\cal R}\wedge^{*} {\cal R} =  24\int e^{\hat 0}e^{\hat 1}e^{\hat 2}e^{\hat 3} = 24\textrm{Vol}_{\textrm{AdS}_4}~.
\end{equation}
For global AdS$_4$ with metric (\ref{eq:globads}) we regulate the volume by a surface at some constant value radial $\rho_0$ and find
\begin{eqnarray}
\textrm{Vol}_{\textrm{AdS}_4} = 2\pi^2  \int_0^{\rho_0} d\rho \sinh^3\rho =  2\pi^2  \left( {1\over 3}\cosh\rho_0 (\sinh^2\rho_0 -2)  + {2\over 3}\right)~.
\end{eqnarray}
Recall that we take $\ell_A =1$ at this point of the paper. The boundary term added when considering the Gauss-Bonnet theorem with a boundary is \cite{Eguchi:1980jx}
\begin{equation}
- 2\int \epsilon_{abcd}\theta^a_{~b}{\cal R}^c_{~d} + {4\over 3}\int\epsilon_{abcd}\theta^a_{~b}\theta^c_{~e}\theta^e_{~d}
= - 24\cdot {1\over 3} \cosh\rho_0 (\sinh^2\rho_0 -2)2\pi^2~,
\end{equation}
where the second fundamental form $\theta^a_{~b}$ is essentially the connection $1$-form and has nonvanishing components
\begin{equation}
\theta_{{\hat\rho}{\hat i}}= \omega_{{\hat\rho}{\hat i}} = - {\cosh\rho\over\sinh\rho} e^{\hat i}~.
\end{equation}
The sum of the bulk and boundary terms then gives
\begin{align}
\chi
&= {1\over 32\pi^2}\cdot 24\cdot {2\over 3}\cdot 2\pi^2 =  1~,
\end{align}
after including the correct overall numerical factor already quoted in (\ref{eq:gaussbonnet}). The cancellation of the terms that diverge at large $\rho_0$ is guaranteed by topological invariance and the role of the boundary terms is to make this happen. The finite term that remains is essentially the regularized volume of AdS$_4$, except for the constant factor $E_4=24$.

The important point is that AdS$_4$ with $S^1\times S^2$ boundary works out qualitatively differently. The metric is thermal AdS$_4$
\begin{equation}
ds^2_4 = \cosh^2\rho d\tau^2 + d\rho^2 + \sinh^2\rho d\Omega^2_2~.
\end{equation}
Taking the circumference of $S^1$ to be $\beta$, the bulk term (\ref{eq:bulke4}) with a regulator in the new radial coordinate $\rho$ gives 
\begin{equation}
24\textrm{Vol}_{\textrm{AdS}_4} = 24\int_0^{\rho_0} \cosh\rho\sinh^2\rho d\rho\cdot\beta\cdot 4\pi 
= 32\pi\beta\sinh^3\rho_0~,
\end{equation}
and the boundary term is
\begin{equation}
-4\int\theta_{{\hat\rho}{\hat i}} R_{{\hat j}{\hat k}}\epsilon^{{\hat\rho}{\hat i}{\hat j}{\hat k}}  +
{4\over 3}\int\epsilon_{abcd}\theta^a_{~b}\theta^c_{~e}\theta^e_{~d} =  -8 \sinh^3\rho_0 \cdot 4\pi\beta~.
\end{equation}
The sum vanishes,
\begin{equation}
\chi = 0~.
\end{equation}

The difference in topology is significant because the divergence and the corresponding physical logarithm depends on topology. We primarily study global AdS$_4$ with $S^3$ boundary conditions because for $\chi=1$ there is a divergence. In thermal AdS$_4$ the boundary is $S^1\times S^2$ and the $S^1$ guarantees a discrete spectrum. This gives technical simplifications but it also excludes the divergence altogether since then $\chi=0$. 

Quantum inequivalence between antisymmetric tensors and scalar fields also depends on the Euler 
number $\chi$ so similar comments apply. In AdS$_4$ with $S^3$ boundary conditions there is quantum inequivalence which we interpret as due to boundary modes. In AdS$_4$ with $S^1\times S^2$ boundary there is quantum equivalence and no boundary modes. Thus it appears that there is a precise sense in which the number of boundary modes is $n^{\rm bndy}=\chi$ despite the subtleties due to noncompactness of AdS$_4$.  


\acknowledgments

We are grateful to Marcos Marino and Ashoke Sen for discussions and early collaboration on this project. 
We also thank Zvi Bern, Massimo Bianchi, Bernard deWit, Mike Duff, Henriette Elvang, Simone Giombi, 
Neil Lambert, Jim Liu, Eric Perlmutter, and Kelly Stelle for discussions and encouragement.

This work was supported in part by the U.S. Department of Energy under grant DE-FG02-95ER40899.


\bibliography{ads4_bibliography}

\providecommand{\href}[2]{#2}\begingroup\raggedright\begin{thebibliography}{10}

\bibitem{Bern:2006kd}
Z.~Bern, L.~J. Dixon, and R.~Roiban, {\it {Is N = 8 supergravity ultraviolet
  finite?}},  {\em Phys. Lett.} {\bf B644} (2007) 265--271,
  [\href{http://arxiv.org/abs/hep-th/0611086}{{\tt hep-th/0611086}}].

\bibitem{Bern:2007xj}
Z.~Bern, J.~J. Carrasco, D.~Forde, H.~Ita, and H.~Johansson, {\it {Unexpected
  Cancellations in Gravity Theories}},  {\em Phys. Rev.} {\bf D77} (2008)
  025010, [\href{http://arxiv.org/abs/0707.1035}{{\tt arXiv:0707.1035}}].

\bibitem{Kallosh:2008mq}
R.~Kallosh, {\it {On a possibility of a UV finite N=8 supergravity}},
  \href{http://arxiv.org/abs/0808.2310}{{\tt arXiv:0808.2310}}.

\bibitem{BjerrumBohr:2009zz}
N.~E.~J. Bjerrum-Bohr and P.~Vanhove, {\it {Surprising simplicity of N=8
  supergravity}},  {\em Int. J. Mod. Phys.} {\bf D18} (2009) 2295--2301.

\bibitem{Kallosh:2009db}
R.~Kallosh, {\it {N=8 Supergravity on the Light Cone}},  {\em Phys. Rev.} {\bf
  D80} (2009) 105022, [\href{http://arxiv.org/abs/0903.4630}{{\tt
  arXiv:0903.4630}}].

\bibitem{Bjornsson:2010wm}
J.~Bjornsson and M.~B. Green, {\it {5 loops in 24/5 dimensions}},  {\em JHEP}
  {\bf 08} (2010) 132, [\href{http://arxiv.org/abs/1004.2692}{{\tt
  arXiv:1004.2692}}].

\bibitem{Dixon:2010gz}
L.~J. Dixon, {\it {Ultraviolet Behavior of $\mathcal{N}=8$ Supergravity}},
  {\em Subnucl. Ser.} {\bf 47} (2011) 1--39,
  [\href{http://arxiv.org/abs/1005.2703}{{\tt arXiv:1005.2703}}].

\bibitem{Christensen:1978gi}
S.~Christensen and M.~Duff, {\it {Axial and Conformal Anomalies for Arbitrary
  Spin in Gravity and Supergravity}},  {\em Phys.Lett.} {\bf B76} (1978) 571.

\bibitem{Christensen:1978md}
S.~Christensen and M.~Duff, {\it {New Gravitational Index Theorems and
  Supertheorems}},  {\em Nucl.Phys.} {\bf B154} (1979) 301.

\bibitem{Christensen:1979iy}
S.~Christensen and M.~Duff, {\it {Quantizing Gravity with a Cosmological
  Constant}},  {\em Nucl.Phys.} {\bf B170} (1980) 480.

\bibitem{Sezgin:1980tp}
E.~Sezgin and P.~van Nieuwenhuizen, {\it {Renormalizability Properties of
  Antisymmetric Tensor Fields Coupled to Gravity}},  {\em Phys. Rev.} {\bf D22}
  (1980) 301.

\bibitem{Bastianelli:2005vk}
F.~Bastianelli, P.~Benincasa, and S.~Giombi, {\it {Worldline approach to vector
  and antisymmetric tensor fields}},  {\em JHEP} {\bf 04} (2005) 010,
  [\href{http://arxiv.org/abs/hep-th/0503155}{{\tt hep-th/0503155}}].

\bibitem{Bastianelli:2005uy}
F.~Bastianelli, P.~Benincasa, and S.~Giombi, {\it {Worldline approach to vector
  and antisymmetric tensor fields. II.}},  {\em JHEP} {\bf 10} (2005) 114,
  [\href{http://arxiv.org/abs/hep-th/0510010}{{\tt hep-th/0510010}}].

\bibitem{Siegel:1980jj}
W.~Siegel, {\it {Hidden Ghosts}},  {\em Phys. Lett.} {\bf B93} (1980) 170.

\bibitem{Duff:1980qv}
M.~J. Duff and P.~van Nieuwenhuizen, {\it {Quantum Inequivalence of Different
  Field Representations}},  {\em Phys. Lett.} {\bf B94} (1980) 179.

\bibitem{Siegel:1980ax}
W.~Siegel, {\it {Quantum Equivalence of Different Field Representations}},
  {\em Phys. Lett.} {\bf B103} (1981) 107.

\bibitem{Gibbons:1984dg}
G.~W. Gibbons and H.~Nicolai, {\it {One Loop Effects on the Round Seven
  Sphere}},  {\em Phys. Lett.} {\bf B143} (1984) 108--114.

\bibitem{Inami:1984vp}
T.~Inami and K.~Yamagishi, {\it {Vanishing Quantum Vacuum Energy in
  Eleven-dimensional Supergravity on the Round Seven Sphere}},  {\em Phys.
  Lett.} {\bf B143} (1984) 115--120.

\bibitem{Fradkin:1984ai}
E.~S. Fradkin and A.~A. Tseytlin, {\it {Quantum Equivalence of Dual Field
  Theories}},  {\em Annals Phys.} {\bf 162} (1985) 31.

\bibitem{Buchbinder:2008jf}
I.~L. Buchbinder, E.~N. Kirillova, and N.~G. Pletnev, {\it {Quantum Equivalence
  of Massive Antisymmetric Tensor Field Models in Curved Space}},  {\em Phys.
  Rev.} {\bf D78} (2008) 084024, [\href{http://arxiv.org/abs/0806.3505}{{\tt
  arXiv:0806.3505}}].

\bibitem{Maldacena:1997re}
J.~M. Maldacena, {\it {The Large N limit of superconformal field theories and
  supergravity}},  {\em Int. J. Theor. Phys.} {\bf 38} (1999) 1113--1133,
  [\href{http://arxiv.org/abs/hep-th/9711200}{{\tt hep-th/9711200}}]. [Adv.
  Theor. Math. Phys.2,231(1998)].

\bibitem{Nicolai:1983me}
H.~Nicolai, {\it {INTRODUCTION TO SUPERSYMMETRY AND SUPERGRAVITY}},  {\em Acta
  Phys. Austriaca Suppl.} {\bf 25} (1983) 71--100.

\bibitem{Mansfield:2003gs}
P.~Mansfield, D.~Nolland, and T.~Ueno, {\it {The Boundary Weyl anomaly in the
  N=4 SYM / type IIB supergravity correspondence}},  {\em JHEP} {\bf 01} (2004)
  013, [\href{http://arxiv.org/abs/hep-th/0311021}{{\tt hep-th/0311021}}].

\bibitem{Beccaria:2014xda}
M.~Beccaria and A.~A. Tseytlin, {\it {Higher spins in AdS$\_{5}$ at one loop:
  vacuum energy, boundary conformal anomalies and AdS/CFT}},  {\em JHEP} {\bf
  11} (2014) 114, [\href{http://arxiv.org/abs/1410.3273}{{\tt
  arXiv:1410.3273}}].

\bibitem{Giombi:2013fka}
S.~Giombi and I.~R. Klebanov, {\it {One Loop Tests of Higher Spin AdS/CFT}},
  {\em JHEP} {\bf 1312} (2013) 068, [\href{http://arxiv.org/abs/1308.2337}{{\tt
  arXiv:1308.2337}}].

\bibitem{Giombi:2014iua}
S.~Giombi, I.~R. Klebanov, and B.~R. Safdi, {\it {Higher Spin
  AdS$\_{d+1}$/CFT$\_d$ at One Loop}},  {\em Phys. Rev.} {\bf D89} (2014),
  no.~8 084004, [\href{http://arxiv.org/abs/1401.0825}{{\tt arXiv:1401.0825}}].

\bibitem{Giombi:2014yra}
S.~Giombi, I.~R. Klebanov, and A.~A. Tseytlin, {\it {Partition Functions and
  Casimir Energies in Higher Spin AdS\_{d+1}/CFT\_d}},  {\em Phys. Rev.} {\bf
  D90} (2014), no.~2 024048, [\href{http://arxiv.org/abs/1402.5396}{{\tt
  arXiv:1402.5396}}].

\bibitem{Banerjee:2010qc}
S.~Banerjee, R.~K. Gupta, and A.~Sen, {\it {Logarithmic Corrections to Extremal
  Black Hole Entropy from Quantum Entropy Function}},  {\em JHEP} {\bf 1103}
  (2011) 147, [\href{http://arxiv.org/abs/1005.3044}{{\tt arXiv:1005.3044}}].

\bibitem{Banerjee:2011jp}
S.~Banerjee, R.~K. Gupta, I.~Mandal, and A.~Sen, {\it {Logarithmic Corrections
  to $\mathcal{N}=4$ and $\mathcal{N}=8$ Black Hole Entropy: A One Loop Test of
  Quantum Gravity}},  {\em JHEP} {\bf 1111} (2011) 143,
  [\href{http://arxiv.org/abs/1106.0080}{{\tt arXiv:1106.0080}}].

\bibitem{Sen:2011ba}
A.~Sen, {\it {Logarithmic Corrections to $\mathcal{N}=2$ Black Hole Entropy: An
  Infrared Window into the Microstates}},
  \href{http://arxiv.org/abs/1108.3842}{{\tt arXiv:1108.3842}}.

\bibitem{Sen:2012cj}
A.~Sen, {\it {Logarithmic Corrections to Rotating Extremal Black Hole Entropy
  in Four and Five Dimensions}},  {\em Gen.Rel.Grav.} {\bf 44} (2012)
  1947--1991, [\href{http://arxiv.org/abs/1109.3706}{{\tt arXiv:1109.3706}}].

\bibitem{Bhattacharyya:2012wz}
S.~Bhattacharyya, B.~Panda, and A.~Sen, {\it {Heat Kernel Expansion and
  Extremal Kerr-Newmann Black Hole Entropy in Einstein-Maxwell Theory}},  {\em
  JHEP} {\bf 1208} (2012) 084, [\href{http://arxiv.org/abs/1204.4061}{{\tt
  arXiv:1204.4061}}].

\bibitem{Sen:2012dw}
A.~Sen, {\it {Logarithmic Corrections to Schwarzschild and Other Non-extremal
  Black Hole Entropy in Different Dimensions}},  {\em JHEP} {\bf 1304} (2013)
  156, [\href{http://arxiv.org/abs/1205.0971}{{\tt arXiv:1205.0971}}].

\bibitem{Sen:2014aja}
A.~Sen, {\it {Microscopic and Macroscopic Entropy of Extremal Black Holes in
  String Theory}},  {\em Gen.Rel.Grav.} {\bf 46} (2014) 1711,
  [\href{http://arxiv.org/abs/1402.0109}{{\tt arXiv:1402.0109}}].

\bibitem{Keeler:2014bra}
C.~Keeler, F.~Larsen, and P.~Lisbao, {\it {Logarithmic Corrections to
  $\mathcal{N} \geq 2$ Black Hole Entropy}},  {\em Phys.Rev.} {\bf D90} (2014),
  no.~4 043011, [\href{http://arxiv.org/abs/1404.1379}{{\tt arXiv:1404.1379}}].

\bibitem{Larsen:2014bqa}
F.~Larsen and P.~Lisbao, {\it {Quantum Corrections to Supergravity on
  AdS$\_2\times S^2$}},  {\em Phys. Rev.} {\bf D91} (2015), no.~8 084056,
  [\href{http://arxiv.org/abs/1411.7423}{{\tt arXiv:1411.7423}}].

\bibitem{Eguchi:1980jx}
T.~Eguchi, P.~B. Gilkey, and A.~J. Hanson, {\it {Gravitation, Gauge Theories
  and Differential Geometry}},  {\em Phys. Rept.} {\bf 66} (1980) 213.

\bibitem{Birrell:1982}
N.~D. Birrell and P.~C.~W. Davies, {\em {Quantum Fields in Curved Space}}.
\newblock Cambridge University Press, Cambridge, UK, 1982.

\bibitem{Vassilevich:2003xt}
D.~Vassilevich, {\it {Heat kernel expansion: User's manual}},  {\em Phys.Rept.}
  {\bf 388} (2003) 279--360, [\href{http://arxiv.org/abs/hep-th/0306138}{{\tt
  hep-th/0306138}}].

\bibitem{Camporesi:1992wn}
R.~Camporesi and A.~Higuchi, {\it {Stress energy tensors in anti-de Sitter
  space-time}},  {\em Phys. Rev.} {\bf D45} (1992) 3591--3603.

\bibitem{Camporesi:1994ga}
R.~Camporesi and A.~Higuchi, {\it {Spectral functions and zeta functions in
  hyperbolic spaces}},  {\em J. Math. Phys.} {\bf 35} (1994) 4217--4246.

\bibitem{Higuchi:1986wu}
A.~Higuchi, {\it {Symmetric Tensor Spherical Harmonics on the $N$ Sphere and
  Their Application to the De Sitter Group SO($N$,1)}},  {\em J. Math. Phys.}
  {\bf 28} (1987) 1553. [Erratum: J. Math. Phys.43,6385(2002)].

\bibitem{Metsaev:1994ys}
R.~R. Metsaev, {\it {Lowest eigenvalues of the energy operator for totally
  (anti)symmetric massless fields of the n-dimensional anti-de Sitter group}},
  {\em Class. Quant. Grav.} {\bf 11} (1994) L141--L145.

\bibitem{Camporesi:1993mz}
R.~Camporesi and A.~Higuchi, {\it {Arbitrary spin effective potentials in
  anti-de Sitter space-time}},  {\em Phys. Rev.} {\bf D47} (1993) 3339--3344.

\bibitem{Taylor:1989ua}
T.~R. Taylor and G.~Veneziano, {\it {Quantum Gravity at Large Distances and the
  Cosmological Constant}},  {\em Nucl. Phys.} {\bf B345} (1990) 210--230.

\bibitem{Bytsenko:1994at}
A.~A. Bytsenko, S.~D. Odintsov, and S.~Zerbini, {\it {The Effective action in
  gauged supergravity on hyperbolic background and induced cosmological
  constant}},  {\em Phys. Lett.} {\bf B336} (1994) 355--361,
  [\href{http://arxiv.org/abs/hep-th/9408095}{{\tt hep-th/9408095}}].

\bibitem{Freedman:1976aw}
D.~Z. Freedman and A.~K. Das, {\it {Gauge Internal Symmetry in Extended
  Supergravity}},  {\em Nucl. Phys.} {\bf B120} (1977) 221.

\bibitem{deWit:1981yv}
B.~de~Wit and H.~Nicolai, {\it {Extended Supergravity With Local SO(5)
  Invariance}},  {\em Nucl. Phys.} {\bf B188} (1981) 98.

\bibitem{Grisaru:1984vk}
M.~T. Grisaru, N.~K. Nielsen, W.~Siegel, and D.~Zanon, {\it {Energy Momentum
  Tensors, Supercurrents, (Super)traces and Quantum Equivalence}},  {\em Nucl.
  Phys.} {\bf B247} (1984) 157.

\bibitem{Green:2007zzb}
M.~B. Green, H.~Ooguri, and J.~H. Schwarz, {\it {Nondecoupling of Maximal
  Supergravity from the Superstring}},  {\em Phys. Rev. Lett.} {\bf 99} (2007)
  041601, [\href{http://arxiv.org/abs/0704.0777}{{\tt arXiv:0704.0777}}].

\bibitem{Bhattacharyya:2012ye}
S.~Bhattacharyya, A.~Grassi, M.~Marino, and A.~Sen, {\it {A One-Loop Test of
  Quantum Supergravity}},  {\em Class.Quant.Grav.} {\bf 31} (2014) 015012,
  [\href{http://arxiv.org/abs/1210.6057}{{\tt arXiv:1210.6057}}].

\bibitem{Marino:2009jd}
M.~Marino and P.~Putrov, {\it {Exact Results in ABJM Theory from Topological
  Strings}},  {\em JHEP} {\bf 06} (2010) 011,
  [\href{http://arxiv.org/abs/0912.3074}{{\tt arXiv:0912.3074}}].

\bibitem{Fuji:2011km}
H.~Fuji, S.~Hirano, and S.~Moriyama, {\it {Summing Up All Genus Free Energy of
  ABJM Matrix Model}},  {\em JHEP} {\bf 08} (2011) 001,
  [\href{http://arxiv.org/abs/1106.4631}{{\tt arXiv:1106.4631}}].

\bibitem{Marino:2011eh}
M.~Marino and P.~Putrov, {\it {ABJM theory as a Fermi gas}},  {\em J. Stat.
  Mech.} {\bf 1203} (2012) P03001, [\href{http://arxiv.org/abs/1110.4066}{{\tt
  arXiv:1110.4066}}].

\bibitem{Camporesi:0000}
R.~Camporesi and A.~Higuchi, {\it {The Plancherel measure for p-forms in real
  hyperbolic spaces}},  {\em J.Geom.Phys.} {\bf 15} (1994).

\bibitem{Avis:1977yn}
S.~J. Avis, C.~J. Isham, and D.~Storey, {\it {Quantum Field Theory in anti-De
  Sitter Space-Time}},  {\em Phys. Rev.} {\bf D18} (1978) 3565.

\bibitem{Breitenlohner:1982jf}
P.~Breitenlohner and D.~Z. Freedman, {\it {Stability in Gauged Extended
  Supergravity}},  {\em Annals Phys.} {\bf 144} (1982) 249.

\end{thebibliography}\endgroup
\bibliographystyle{JHEP}

\end{document}